\documentclass[11pt,a4paper]{article}
\pdfoutput=1
\usepackage{jheppub}
\usepackage{amsfonts,amsthm}
\usepackage{amsmath,amssymb}
\usepackage{graphicx,color}
\usepackage[utf8]{inputenc}
\newcommand{\be}{\begin{equation}}
\newcommand{\ee}{\end{equation}}
\newcommand{\bea}{\begin{eqnarray}}
\newcommand{\eea}{\end{eqnarray}}

\renewcommand{\Re}{\mathrm{Re }}

\newcommand{\doublet}[2]{ \left( \begin{array}{c} {#1} \\ {#2} \end{array}\right) }

\newcommand{\lr}[1]{ \langle #1 \rangle}

\newcommand{\Z}{\mathbb{Z}}
\newcommand{\mmatrix}[4]{ \left(\! \begin{array}{ccc}#1 & #2 \\ #3 & #4 \end{array}\!\right) }
\newcommand{\mmmatrix}[9]{ \left(\! \begin{array}{ccc}#1 & #2 & #3\\ #4 & #5 & #6\\ #7 & #8 & #9\\ \end{array}\!\right) }
\newcommand{\toCP}{\xrightarrow{CP}}
\newcommand{\lrpartial}{\,\partial^{\hspace{-7pt}\raise3pt\hbox{\small $\leftrightarrow$}}\!}

\def\lsim{\mathrel{\rlap{\lower4pt\hbox{\hskip1pt$\sim$}}
    \raise1pt\hbox{$<$}}}         
\def\gsim{\mathrel{\rlap{\lower4pt\hbox{\hskip1pt$\sim$}}
    \raise1pt\hbox{$>$}}}         

\title{Radiative neutrino masses from order-4 $CP$ symmetry}

\author{Igor P. Ivanov}

\affiliation{CFTP, Instituto Superior T\'ecnico, Universidade de Lisboa, av. Rovisco Pais 1, 1049-001 Lisboa, Portugal}
\emailAdd{igor.ivanov@tecnico.ulisboa.pt}

\abstract{
Generalized $CP$ symmetry of order 4 (CP4) is surprisingly powerful in shaping scalar and quark sectors 
of multi-Higgs models. Here, we extend this framework to the neutrino sector.
We build two simple Majorana neutrino mass models with unbroken CP4, which are analogous to Ma's scotogenic model.
Both models use three Higgs doublets and two or three right-handed (RH) neutrinos.
The minimal CP4 symmetric scotogenic model uses only two RH neutrinos, 
leads to three non-zero light neutrino masses, and contains a built-in mechanism 
to further suppress them via phase alignment.
With three RH neutrinos, one generates a type I seesaw mass matrix of rank 1,
which is then corrected by the same scotogenic mechanism, 
naturally leading to two neutrino mass scales with mild hierarchy.
These minimal CP4-based constructions emerge as a primer for introducing additional symmetry structures
and exploring their phenomenological consequences.
}

\begin{document}
\maketitle

\section{Introduction}

The tiny mass scale of neutrinos and their mixing patterns are considered by many 
a clear indication that a new mechanism beyond the Standard Model (bSM) is at work.
Dozens of neutrino mass models with different levels of sophistication 
have been proposed, and many of them are based on symmetry arguments, 
see reviews \cite{King:2017guk,Cai:2017jrq,King:2014nza,Altarelli:2010gt} and references therein.
Some models aim to quantitatively reproduce the mass and mixing parameters and 
employ for that purpose discrete or continuous symmetry groups and various new field multiplets
transforming non-trivially under them.
Others keep assumptions to the minimum and propose new qualitatively different mechanisms
for neutrino mass generation. 

One appealing example of the latter class of models is the scotogenic model 
suggested by Ma in 2006 \cite{Ma:2006km}. It makes use of an additional ``inert'' Higgs doublet
and the right-handed neutrinos, which are assumed to be odd under the new global $\Z_2$-symmetry.
If $\Z_2$ remains unbroken, the traditional tree-level seesaw mechanism is not at work.
However, at one loop, the new inert scalars including the scalar dark matter (DM) candidate,
generate the light neutrino mass matrix.
Apart from rich DM consequences, this model may be testable at the LHC \cite{Ho:2013hia,Hessler:2016kwm} or in
lepton-flavor violating (LFV) processes \cite{Toma:2013zsa,Vicente:2014wga}.

In addition to proposing a radiative neutrino mass model, 
Ma's 2006 paper \cite{Ma:2006km} together with Refs.~\cite{Barbieri:2006dq,LopezHonorez:2006gr} 
boosted the exploration of its scalar sector known as the inert doublet model, see, for example, the recent review \cite{Ivanov:2017dad}.
Various more elaborate scalar sectors with DM candidates have been studied later \cite{Keus:2014jha,Bonilla:2014xba,Cordero-Cid:2016krd}.
Such models can also be used to radiatively generate neutrino masses;
the classification of one-loop \cite{Restrepo:2013aga} and
two-loop \cite{Simoes:2017kqb} neutrino mass models generated by scalar DM candidates were established.
In those models, one usually keeps $\Z_2$ as the symmetry that protects DM candidates although 
other family symmetries and their use for radiative neutrino mass generation have also been studied \cite{Farzan:2012ev}.

The intrinsic weakness of $\Z_2$ or $\Z_N$ family symmetries in multi-scalar models
is that they still allow for many free parameters.
Recently, a rather special multi-Higgs-doublet sector was proposed, 
in which the scalar DM candidates are protected by a $CP$ symmetry \cite{Ivanov:2015mwl}.
Unlike all previously constructed models, this model used a generalized $CP$ symmetry of order 4 (called CP4),
which means that one must apply it four times to get the identity transformation.
Imposing CP4 without producing additional accidental symmetry requires three Higgs doublets;
this possibility was found in the course of systematic search for all symmetry groups 
available in three-Higgs-doublet models (3HDM) \cite{Ivanov:2011ae}.

Imposing CP4 leads to a rather well shaped scalar sector \cite{Ivanov:2015mwl}. 
If CP4 remains unbroken, it produces pairwise mass-degenerate scalar sector and DM candidates with peculiar properties.
The same CP4 can also be extended to the quark sector \cite{Aranda:2016qmp,Ferreira:2017tvy},
also strongly shaping the Yukawa matrices. However, CP4 must be broken in this case to avoid mass-degenerate
fermions.

These results naturally lead to the question of whether CP4 can be extended to the neutrino sector
and, remaining unbroken, can produce a new, CP4-based version of the scotogenic model.
In this paper we answer these questions in the affirmative. 
We first construct a CP4-symmetric neutrino sector and then build two minimalistic models
with unbroken CP4, in which DM candidates play the key role in generating light neutrino masses.
Despite the number of free parameters increases when we pass from two to three Higgs doublets,
CP4 alone constrains these models {\em stronger} than $\Z_2$ in the original scotogenic model
and generates features which were absent there. 

The structure of the paper is the following.
In the next section we give essential details of the two main ingredients: the original Ma's scotogenic model
and the CP4-symmetric 3HDM. Then, in section~\ref{section-main}, we extend CP4 to the neutrino sector,
and the consider two minimal examples with two and with three RH neutrinos.
We wrap up the paper with conclusions.

\section{Scalar DM candidates and radiative neutrino masses}

\subsection{Ma's scotogenic model}

We begin with a recap of Ma's scotogenic model proposed in \cite{Ma:2006km}. 
The model postulates a new global symmetry $\Z_2$, under which the SM fields are even,
and adds a second electroweak Higgs doublet $\Phi_2$ 
and three RH neutrinos\footnote{Strictly speaking, since these RH singlets $N_i$ 
do not mix, in the scotogenic model, with the LH neutrinos, naming them RH ``neutrinos''
is a slight abuse of notation. But since this is a widespread terminology,
we will stick to it.} $N_i$, all of which are odd.
In this way the charged leptons are coupled only to the SM-like doublet $\Phi_1$, 
but the Yukawa neutrino interactions mediated by $\Phi_2$ as well as 
the Majorana mass term for $N$'s are allowed:
\be
-{\cal L}_{\mathrm{lept.}} = \Gamma_{\alpha \beta} \overline{L_\alpha} \Phi_1 \ell_{R\, \beta}
+ Y_{\alpha k} \overline{L_\alpha} \tilde \Phi_2 N_k 
+ {1 \over 2} M_{ij} \overline{N^c_i} N_j + h.c.\label{nu-lagrangian-ma}
\ee
Here, the Greek letters $\alpha, \beta = 1, 2, 3$ denote charged leptons and 
the Roman letters $i, j, k = 1, 2, 3$ denote RH neutrinos. 
The $\Z_2$ symmetry stays unbroken upon minimization of the Higgs potential,
$\lr{\Phi_2} = 0$,
which can be easily achieved in a significant part of the scalar sector parameter space.
The second, inert doublet $\Phi_2$ produces two neutral scalars denoted traditionally as $H$ and $A$
plus a pair of charged Higgses $H^\pm$, and the lightest among them, usually taken to be $H$,
is the DM candidate (unless some of the new heavy neutrinos is even lighter).
Due to $\lr{\Phi_2} = 0$, no Dirac mass term appears, and the LH neutrinos remain massless at the tree level.

\begin{figure}[h]
\begin{center}
\includegraphics[width=6cm]{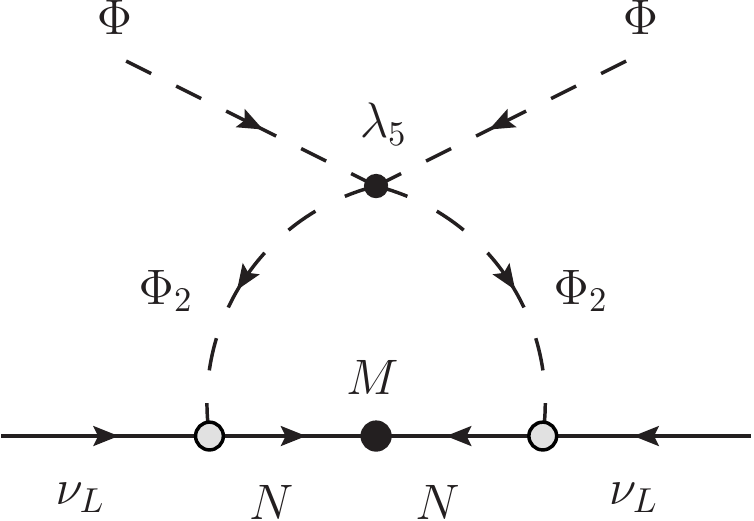}\hspace{1cm}
\includegraphics[width=5.5cm]{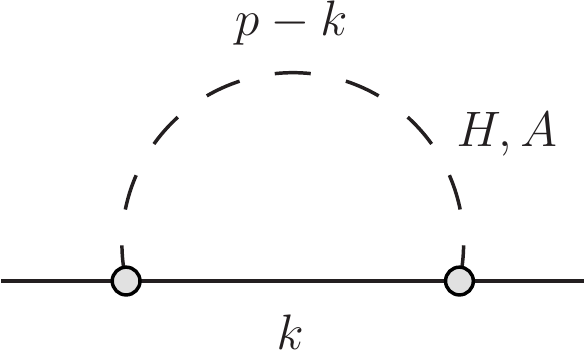}
\caption{Left: the standard diagram representing the radiative neutrino mass generation in the 
scotogenic model. Right: the actual $H$ and $A$-mediated loop diagrams one needs to calculate.}
\label{fig1}
\end{center}
\end{figure}

However, these masses are generated at the 1-loop level via loop diagrams mediated by the inert neutral scalars $H$ and $A$,
Fig.~\ref{fig1}.
Since the fermion line involves scalar interactions of the form
\be
\cdots Y_{\alpha k} {1 \over \sqrt{2}}(H - i A) \cdots Y_{\beta k} {1 \over \sqrt{2}}(H - i A) \cdots\,,
\ee
one gets the difference between the $H$-loop and the $A$-loop with different masses $m_H \equiv m$ and $m_A \equiv M$.
The scalar potential contains, among other, the interaction term $\lambda_5 [(\Phi_1^\dagger \Phi_2)^2 + h.c.]/2$,
which generates the $H/A$ mass splitting:
\be
m_H^2 - m_A^2 = \lambda_5 v^2\,. 
\ee
Thus, the two 1-loop diagrams cancel only in their divergent parts and produce the finite piece
proportional to the following scalar function:
\be
J(m,M; M_k) = {M_k \over 16 \pi^2}
\left({m^2 \over M_k^2 - m^2}\log {M_k^2 \over m^2} - {M^2 \over M_k^2 - M^2}\log {M_k^2 \over M^2} \right)\,.\label{loop-J}
\ee
The resulting light neutrino mass matrix is written as 
\be
({\cal M}_\nu)_{\alpha\beta} = {1 \over 2}\sum_k Y_{\alpha k} Y_{\beta k} \cdot J(m,M; M_k)\,.\label{mass-matrix-ma}
\ee
In particular, for small mass splitting and heavy RH neutrinos, $\lambda_5 v^2 \ll m_0^2 \equiv (m^2 + M^2)/2 \ll M_k^2$, 
the scalar function $J$ can be simplified as
\be
J \approx {\lambda_5 v^2 \over 16 \pi^2 M_k} \left(\log {M_k^2 \over m_0^2} - 1\right)\,.
\ee
Thus, with respect to the classical seesaw mechanism, the scotogenic model
generates an extra suppression factor $\lambda_5 /(32\pi^2)$ potentially enhanced by the logarithm, 
and it can be rather small.

The minimum of assumptions is a very appealing feature of the model.
One just adds an extra doublet, (usually) three RH neutrinos, and the smallest finite group $\Z_2$,
and naturally derives several qualitative consequences.
On the other hand, although it can provide tiny neutrino masses for reasonably large $M_k$, 
it cannot predict patterns in the mass matrix ${\cal M}_\nu$, since the Yukawa couplings $Y_{\alpha k}$ can be arbitrary.

\subsection{CP4 3HDM}

We aim to apply this scotogenic idea not to the inert doublet model but to the CP4 3HDM, 
three-Higgs-doublet model model equipped with the generalized $CP$ symmetry of order 4 (CP4)
and no other accidental symmetries.
It was very briefly mentioned in the last appendix of \cite{Ivanov:2011ae},
then brought up in \cite{Ivanov:2015mwl} as an example of a $CP$-conserving model without real basis,
and later studied in finer detail in \cite{Aranda:2016qmp}.
Without loss of generality, its scalar potential can be written as $V = V_0 + V_1$, where
\bea
V_0 &=& - m_{11}^2 (\Phi_1^\dagger \Phi_1) - m_{22}^2 \left(\Phi_2^\dagger \Phi_2 + \Phi_3^\dagger \Phi_3\right)
+ \lambda_1 (\Phi_1^\dagger \Phi_1)^2 + \lambda_2 \left[(\Phi_2^\dagger \Phi_2)^2 + (\Phi_3^\dagger \Phi_3)^2\right]
\nonumber\\
&+& \lambda_3 (\Phi_1^\dagger \Phi_1) \left(\Phi_2^\dagger \Phi_2 + \Phi_3^\dagger \Phi_3\right)
+ \lambda'_3 (\Phi_2^\dagger \Phi_2) (\Phi_3^\dagger \Phi_3)\nonumber\\
&+& \lambda_4 \left[(\Phi_1^\dagger \Phi_2)(\Phi_2^\dagger \Phi_1) + (\Phi_1^\dagger \Phi_3)(\Phi_3^\dagger \Phi_1)\right]
+ \lambda'_4 (\Phi_2^\dagger \Phi_3)(\Phi_3^\dagger \Phi_2) \,,\label{V0}
\eea
and
\be
V_1 = {\lambda_6 \over 2} \left[(\Phi_2^\dagger\Phi_1)^2 - (\Phi_1^\dagger\Phi_3)^2\right] +
\lambda_8(\Phi_2^\dagger \Phi_3)^2 + \lambda_9(\Phi_2^\dagger\Phi_3)\left(\Phi_2^\dagger\Phi_2-\Phi_3^\dagger\Phi_3\right) + h.c.
\label{V1b}
\ee
with real $\lambda_{6}$ but still complex $\lambda_{8,9}$. 
This potential is invariant under
\be
J: \ \phi_i \mapsto X_{ij} \phi_j^*\,, \quad 
X =  \left(\begin{array}{ccc}
1 & 0 & 0 \\
0 & 0 & i \\
0 & -i & 0 
\end{array}\right)\,.
\label{CP4}
\ee
For generic values of the parameters, there are no other global symmetries of this potential. 

Notice that $J^2 = X X^* = \mathrm{diag}(1,\,-1,\,-1)$, so that, as a byproduct, this model effectively incorporates a $\Z_2$
symmetry and uniquely assigns $\Phi_1$ to be the $\Z_2$-even and $\Phi_2$ and $\Phi_3$ to be the $\Z_2$-odd doublets.
If CP4 symmetry is to be conserved in vacuum, we must require that $\lr{\Phi_2} = \lr{\Phi_3} = 0$,
which can be satisfied in a significant part of the scalar parameter space.
In this way, $\Phi_2$ and $\Phi_3$ become inert: they do not contribute 
to the gauge boson or charged fermion masses.
Expanding the potential near the vacuum as
\be
\Phi_1 = \doublet{G^+}{{1 \over \sqrt{2}}(v + h_{125} + i G^0)}\,,\quad
\Phi_2 = \doublet{H^+_2}{{1 \over \sqrt{2}}(H + i a)}\,, \quad
\Phi_3 = \doublet{H^+_3}{{1 \over \sqrt{2}}(h + i A)}\,, \label{fields-basis}
\ee
we can find that all these scalars are already mass eigenstates with the following masses:
\bea
m^2 \equiv m_{h,a}^2 &=& {1\over 2}v^2 (\lambda_3 +\lambda_4 - \lambda_6) - m_{22}^2\,,\nonumber\\
M^2 \equiv m_{H,A}^2 &=& {1\over 2}v^2 (\lambda_3 +\lambda_4 + \lambda_6) - m_{22}^2 = m^2 + \lambda_6 v^2\,,\label{neutral-inert-masses}\\
m_{H^\pm}^2 \equiv m_{H^\pm_2, H^\pm_3}^2 
&=& {1\over 2}v^2 \lambda_3 - m_{22}^2 = m^2 + {1\over 2}v^2 (\lambda_6 -\lambda_4)\,.\label{inert-masses}
\eea
The inert spectrum is pairwise degenerate, which is a rare instance of the state doubling beyond Kramers degeneracy 
\cite{Aranda:2016qmp}. Indeed, the conserved CP4 acts on the neutral scalars as
\be
H \toCP A\,, \quad A\toCP -H\,, \quad h \toCP -a\,, \quad a\toCP h\,.
\ee
It is this symmetry which protects the lightest inert scalars $h$ and $a$ against decay.
We also see that $\lambda_6$ in this model plays the role of $\lambda_5$ of the inert doublet model:
it governs the mass splitting of the two pairs of inert neutral scalars.

CP4 symmetry can also be extended to the fermion sector \cite{Aranda:2016qmp,Ferreira:2017tvy}. 
Requirement that the Yukawa interactions are CP4 invariant forces this transformation
to mix fermion generations as well, in the way similar to (\ref{CP4}).
If CP4 remains unbroken, it leads to mass-degenerate quarks or leptons, which must be avoided.
One can either assume that CP4 is spontaneously 
broken---this route was taken in \cite{Ferreira:2017tvy}---or decouple fermions from the inert doublets altogether.
In the latter path, we can nevertheless allow RH neutrinos to transform non-trivially under CP4,
which will lead us to the desired scotogenic model. 

\section{Radiative neutrino masses from CP4}\label{section-main}

\subsection{CP4 symmetric neutrino sector}

We want to build a scotogenic model based on the generalized $CP$-symmetry CP4 
rather than the family symmetry $\Z_2$.
We work in the same CP4 3HDM scalars sector as before and build a 3HDM analog of Eq.~(\ref{nu-lagrangian-ma}):
\be
-{\cal L}_{\mathrm{lept.}} = \Gamma^{(a)}_{\alpha \beta} \overline{L_\alpha} \Phi_a \ell_{R\,\beta}
+ Y^{(a)}_{\alpha k} \overline{L_\alpha} \tilde \Phi_a N_k 
+ {1 \over 2} M_{ij} \overline{N^c_i} N_j + h.c.\label{nu-lagrangian}
\ee
For the moment, we do not specify the total number of RH neutrinos $N_k$.
CP4 acts on Higgs doublets according to Eq.~(\ref{CP4})
and on fermions as
\be
\Psi_i \toCP U_{ij} \Psi_j^{cp}\,, \quad \mbox{where} \quad \Psi_j^{cp} = \gamma^0 C (\overline{\Psi_j})^T\,,
\ee
where $\Psi$ generically denotes any type of fermions present in (\ref{nu-lagrangian}).
The matrices $U$ accompanying a generalized $CP$ transformation 
can be different for left doublets and right charged lepton and neutrino singlets,
see \cite{Ferreira:2017tvy} for the similar construction in the quark sector.
To avoid massless or mass-degenerate charged leptons, we must assume that CP4 acts
trivially on $L_\alpha$ and $\ell_{R\alpha}$. 
This leads to $\Gamma^{(2)}_{\alpha \beta} = \Gamma^{(3)}_{\alpha \beta} = 0$ 
and an arbitrary real $\Gamma^{(1)}_{\alpha \beta}$.
In terminology of \cite{Ferreira:2017tvy}, this amounts to selecting case $A$
for charged leptons among four possibilities. 

The RH neutrinos $N_k$ can transform non-trivially under CP4.
Requiring that the lagrangian (\ref{nu-lagrangian}) stays invariant leads to the following set of conditions:
\bea
&&Y^{(a)}_{\alpha i} U_{ik} X^*_{ab} = (Y^{(b)}_{\alpha k})^*\,, \label{condition-1} \\
&&(U^T)_{ii'} M_{i'k'} U_{k'k} = M_{ik}^*\,.\label{condition-2}
\eea
Via an appropriate basis change in the RH neutrino space $N_i$, the matrix $U$ can be brought 
to the block-diagonal form \cite{Ecker:1987qp,Weinberg:1995mt}, with the blocks
being either $1\times 1$ phases or $2\times 2$ matrices of the following type:
\be
\mmatrix{c_\alpha}{s_\alpha}{-s_\alpha}{c_\alpha}\quad \mbox{as in \cite{Ecker:1987qp},}\quad \mbox{or}\quad
\mmatrix{0}{e^{i\alpha}}{e^{-i\alpha}}{0}\quad \mbox{as in \cite{Weinberg:1995mt}.}\label{block}
\ee
Condition (\ref{condition-1}) is of the same type as derived for the quark sector \cite{Ferreira:2017tvy}
under the additional assumption that the left-handed fields are not mixed by CP transformation.
The only non-trivial solution available corresponds to $\alpha=\pi/2$, which can also be checked 
by direct derivation.
Thus, the minimal scotogenic model with CP4 requires {\em two RH neutrinos $N_i$}, which
must be coupled only with the doublets $\Phi_2$ and $\Phi_3$ (thus, $Y^{(1)} = 0$). 

To proceed further, we select the first form of matrix $U$:
\be
U_{ij} = \mmatrix{0}{1}{-1}{0}\,,\label{U-selected}
\ee
and convert (\ref{condition-1}) to $-i Y^{(2)} U = (Y^{(3)})^*$ and $i Y^{(3)} U = (Y^{(2)})^*$. 
These conditions lead to the reciprocal dependence of elements of $Y^{(2)}$ and $Y^{(3)}$.
For example, once all elements of $Y^{(2)}$ are chosen, $Y^{(3)}$ is fully reconstructed:
\be
Y^{(2)} = \left(\!\begin{array}{cc}
y_{11} & y_{12} \\
y_{21} & y_{22} \\
y_{31} & y_{32}
\end{array}\!\right), 
\quad 
Y^{(3)} = \left(\!\begin{array}{cc}
-i y_{12}^* & i y_{11}^* \\
-i y_{22}^* & i y_{21}^* \\
-i y_{32}^* & i y_{31}^*
\end{array}\!\right)\,.\label{matrices-Y}
\ee
Alternatively, one could take the first columns in both $Y^{(2)}$ and $Y^{(3)}$ as free parameters,
and then the second columns would be fully determined.
One can also verify by direct computation that
\be
\sum_k Y^{(3)}_{\alpha k} Y^{(3)}_{\beta k} = - \left(\sum_k Y^{(2)}_{\alpha k} Y^{(2)}_{\beta k}\right)^*\,.\label{relation}
\ee
The same $U$ given by Eq.~(\ref{U-selected}) constrains for the RH Majorana mass matrix $M$:
\be
M = \mmatrix{m_{11}}{i m_{12}}{i m_{12}}{m_{11}^*}\,,
\ee
with complex $m_{11}$ and real $m_{12}$.
This complex symmetric matrix can be always brought, via a transformation $V \in SU(2)$ in the $N_i$ space,
to the diagonal form proportional to the identity matrix:
\be
M = V^T D V\,, \quad \mbox{where}\quad D = \mmatrix{M_0}{0}{0}{M_0}\,.
\ee
Notice also that this transformation does not affect the CP4 matrix $U$: $V^T U V = U$.
This is not surprising: the matrix $U$ can be viewed as defining the skew-symmetric product,
and the transformation group which leaves the skew-symmetric product invariant is known as the symplectic group $Sp(1)$
which is isomorphic to $SU(2)$.

Therefore, in the new basis, one can still parametrize Yukawa couplings $Y^{(2)}$ and $Y^{(3)}$ as in (\ref{matrices-Y})
and just replace $M$ with its diagonal form $D$.
In this way, the Majorana mass matrix for RH neutrinos $N_i$ is diagonal and two real degenerate entries $M_0$.

\subsection{Two RH neutrinos: the minimal CP4 scotogenic model}

We are now ready to write the light neutrino mass matrix in the minimal scotogenic model based on CP4.
The expression resembles closely Eq.~(\ref{mass-matrix-ma}):
\be
{\cal M}^s_{\alpha\beta} = {1 \over 2}\sum_{k} \left[Y^{(2)}_{\alpha k} Y^{(2)}_{\beta k} \cdot J(m, M; M_0) + Y^{(3)}_{\alpha k} Y^{(3)}_{\beta k} 
\cdot J(M, m; M_0)\right]\,,\label{expression-1}
\ee
with the same loop function $J$ as in Eq.~(\ref{loop-J}).
Here we used the pairwise mass degeneracy of the inert neutral scalars (\ref{neutral-inert-masses})
and the mass degeneracy between the two RH neutrinos $N_i$.
Since $J(M, m; M_0) = - J(m, M; M_0)$ and using the property (\ref{relation}), 
we can further simplify it as 
\be
{\cal M}^{s}_{\alpha\beta} = \Re\left[\sum_{k} Y^{(2)}_{\alpha k} Y^{(2)}_{\beta k}\right] \cdot J(m, M; M_0)\,.\label{expression-2}
\ee
This is the final expression for the light neutrino mass matrix within the minimal CP4 scotogenic model.
It closely resembles the original scotogenic model result (\ref{mass-matrix-ma})
and supports the intuitive picture that, for unbroken CP4, the duplicated spectrum of inert scalars 
just add up in their loops. 

However, there are several important differences with respect to Ma's scotogenic model.
First, although we used only {\em two} RH neutrinos, the resulting expression does not force the lightest neutrino to be massless.
Indeed, although the matrix $\sum_{k} Y^{(2)}_{\alpha k} Y^{(2)}_{\beta k}$ is of rank 2 and so is its complex conjugation,
their {\em sum} in Eq.~(\ref{expression-1}) can be of rank 3 since they have distinct eigenvectors.\footnote{As an elementary illustration,
consider the following complex symmetric $2\times 2$ matrix $A$:
\be
A = \mmatrix{1}{i}{i}{-1} = \doublet{1}{i}\cdot (\,1,\, i\,) \,.
\ee
This matrix is of rank 1, but $\Re A$ is a matrix of rank 2.}

Second, since ${\cal M}_{\alpha\beta}$ is manifestly real, it is diagonalized by an orthogonal rotation,
which precludes $CP$-violation in the leptonic sector. This is, of course, to be expected:
by construction, we work with an unbroken CP4, and therefore the model cannot display $CP$-violating effects.

Third, since Eq.~(\ref{expression-2}) explicitly uses taking the real part,
one can envision another potential source of suppression or even cancellation of the neutrino masses: 
a common phase $\pi/4$ in all entries of $Y^{(2)}$. Whether such an alignment, exact or approximate,
can be achieved by additional symmetry arguments is an open question.
If it can, then we have extra suppression without fine-tuning among the values of the Yukawa parameters.
It is remarkable that, in that case, neutrino masses will be vanishing at one loop,
but the corresponding lepton flavor violating processes for charged leptons will persist.
Indeed, they involve one-loop diagrams with charged inert scalars and are proportional 
to $\sum_k Y^{(a)}_{\alpha k} [Y^{(a)}_{\beta k}]^*$, see a detailed analysis \cite{Toma:2013zsa,Vicente:2014wga}. 
This expression is always non-vanishing,
and the two mass-degenerate charged scalars from doublets $\Phi_2$ and $\Phi_3$ interfere constructively. 

\subsection{Three neutrino case: two mass scales with mild hierarchy}

Now we turn to the case with three RH neutrinos $N_k$.
The starting expressions (\ref{nu-lagrangian}) as well as the CP4 symmetry conditions (\ref{condition-1})
and (\ref{condition-2}) still hold, but all matrices including the Yukawa couplings $Y^{(a)}$ and the transformation matrix $U$
are now $3\times 3$ matrices.
Again, using the basis change freedom in the $N_k$ space, we make $U$ block-diagonal
\be
U = \mmmatrix{0}{1}{0}{-1}{0}{0}{0}{0}{1}\,,
\ee
which forces the Yukawa matrices $Y^{(a)}$ to be of the following form:
\be
Y^{(1)} = \left(\!\begin{array}{ccc}
0 & 0 & y_{13}\\
0 & 0 & y_{23}\\
0 & 0 & y_{33}
\end{array}\!\right), 
\quad 
Y^{(2)} = \left(\!\begin{array}{ccc}
y_{11} & y_{12} & 0\\
y_{21} & y_{22} & 0\\
y_{31} & y_{32} & 0
\end{array}\!\right), 
\quad 
Y^{(3)} = \left(\!\begin{array}{ccc}
-i y_{12}^* & i y_{11}^* & 0 \\
-i y_{22}^* & i y_{21}^* & 0 \\
-i y_{32}^* & i y_{31}^* & 0
\end{array}\!\right)\,.\label{matrices-Y-3x3}
\ee
Here, the entries of $Y^{(1)}$ are all real, while the entries of $Y^{(2)}$ and $Y^{(3)}$ 
can be complex.
The Majorana mass matrix for the RH neutrinos becomes, after diagonalization, 
\be
M = \mmmatrix{M_0}{0}{0}{0}{M_0}{0}{0}{0}{M_0'}\,,
\ee
where $M_0'$ does not have to be equal to $M_0$.

The main effect of the third RH neutrino is that it can now couple
to the SM-like Higgs doublet $\Phi_1$ via $Y^{(1)}$. Upon the spontaneous symmetry breaking,
it generate the Dirac mass term with the matrix $m_D = v Y^{(1)}/\sqrt{2}$.
Therefore, the standard Type I seesaw mechanism is at work and leads to the following tree-level
light neutrino mass matrix:
\be
{\cal M}^{\mathrm{seesaw}}_{\alpha\beta} = - m_D M^{-1} m_D^T = - {v^2 \over 2M_0'} y_{3\alpha} y_{3\beta}\,.\label{seesaw}
\ee
Once again, this matrix is purely real, so that no $CP$-violating phases emerge.

\begin{figure}[h]
\begin{center}
\includegraphics[width=6cm]{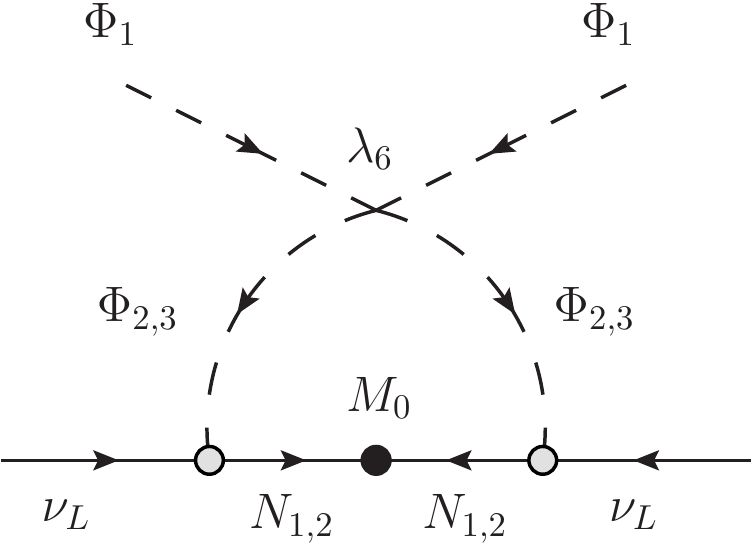}\hspace{1cm}
\includegraphics[width=6cm]{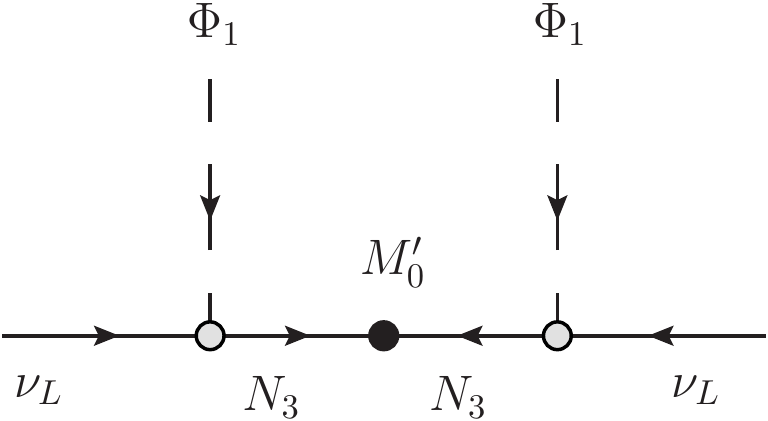}
\caption{In the CP4 symmetric model with three RH neutrinos, 
the first two $N_{1,2}$ generate the scotogenic mass terms, while the third one $N_3$
produces a rank-1 mass term via the usual type 1 seesaw.}
\label{fig2}
\end{center}
\end{figure}

This tree-level mass matrix is not physically acceptable because it is of rank 1 and, therefore, shows insufficient lepton mixing.
But now we recall that the scotogenic mechanism described above with the resulting matrix ${\cal M}^s$ given by 
Eq.~(\ref{expression-2}) is still at work and can be added to the seesaw term:
\be
{\cal M}_\nu = {\cal M}^s + {\cal M}^{\mathrm{seesaw}}\,.\label{final}
\ee
The resulting mechanism and the roles of the three RH neutrinos $N_i$ are schematically shown in Fig.~\ref{fig2}.
Although one cannot predict, on the basis of CP4 alone, the shape of the neutrino mass matrix 
and mixing patterns, there are several qualitative features which emerge from this construction.

First, the full matrix ${\cal M}_\nu$ is of rank 3. However it emerges as a loop-induced correction to 
the tree-level seesaw result, which is unavoidably of rank 1. 
If the loop-induced correction is relatively small, it will naturally generate normally ordered neutrino
masses with two mass scales:
\be
m_1 \sim m_2 \sim {\lambda_6 \over 16 \pi^2} {v^2 \over M_0} [Y^{(2)}]^2 \log\left({M_0^2 \over m^2}\right)\,, \quad 
m_3 \sim {v^2 \over M_0'} [Y^{(1)}]^2\,,
\ee
where $[Y^{(a)}]^2$ schematically denote results of diagonalization of the $Y^{(a)}$ entry products 
which are present in (\ref{expression-2}) and (\ref{seesaw}).

Next, let us we assume, for a rough estimate, that $M_0' = M_0$ and that there is no extra hierarchy 
among the entries of $Y^{(1)}$ and $Y^{(2,3)}$. Then the ratio of the two scales is naturally mild,
\be
{m_{1,2} \over m_3} \sim {\lambda_6 \over 16\pi^2 } \log\left({M_0^2 \over m^2}\right)\,.
\ee
If this ratio is compared with the experimental value of $(\Delta m_{21}^2/|\Delta m_{32}^2|)^{1/2} \approx 0.18$,
one can easily match the two numbers, for any reasonable mass scale $M_0$ and the scalar DM candidate mass $m$,
with $\lambda_6 \sim 1$.
Although we consider this exercise just as a numerical example, 
there may exist a deeper reason for such relations. Namely, if CP4 arises at low-energy 
as the unique residual symmetry of a yet-unknown larger symmetry group with irreducible triplet,
then $M_0' = M_0$ as well as relations among $Y$'s follow automatically.
It is intriguing to construct an explicit example of such a situation.

Of course, no quantitative conclusions can be drawn from this comparison, but at least 
the scales emerging in neutrino mass matrix in the CP4 hybrid seesaw-scotogenic model 
can easily incorporate the experimental data.
The idea that loop contributions can correct the unsatisfactory tree-level neutrino mass matrix 
was exploited as early as in 1999 \cite{Grimus:1999wm} and was recently embedded
in the scotogenic framework in \cite{Hehn:2012kz,Ferreira:2016sbb},
see also other illustrative examples \cite{Hirsch:2004he,Ibarra:2011gn,Wegman:2017zui}.
Our model demonstrates that the same mechanism can be driven by a single albeit non-standard $CP$-symmetry.

\section{Conclusions}

In this work, we saw that the model-building strategy based on a single symmetry CP4,
the generalized $CP$-symmetry of order 4,
which was previously applied to 3HDM scalar \cite{Ivanov:2015mwl} 
and quark sectors \cite{Aranda:2016qmp,Ferreira:2017tvy},
is equally well suited for building neutrino mass models.

To this end, we considered two versions of the model with unbroken CP4 
and with two or three right-handed heavy neutrinos $N_k$.
For two RH neutrinos---which is the minimal case with unbroken CP4---the construction 
resembles Ma's scotogenic model \cite{Ma:2006km} but with additional features driven by CP4.
This minimal model, naturally, does not contain $CP$ violation, allows for three non-zero light neutrino masses
despite using only two RH neutrinos, and has a built-in possibility to further suppress or even cancel this 
mass term via the phase alignment.

For the three RH neutrino case, one predicts the type I seesaw term, which produces the tree-level light neutrino
mass matrix of rank 1. The scotogenic mechanism, which is in action here as well, brings the rank back to 3.
In a way similar to Refs.~\cite{Grimus:1999wm,Hehn:2012kz,Ibarra:2011gn}, 
the model naturally generates two mass scales for neutrino masses, with the relative
magnitude which easily matches the experimentally observed pattern.

Certainly, these minimal models, as they stand, are not in the position to claim quantitative predictions.
Just like the original scotogenic model, 
the power of these CP4-based models is the minimality of their assumptions and surprisingly far-fetched consequences.
We show that using truly minimal models realizing unbroken CP4 without any freedom to assign representations,
we get neutrino masses and potentially rich physics.
Of course, one can elaborate on it by considering spontaneously broken CP4 \cite{Ferreira:2017tvy}
or even allowing for softly CP4-breaking terms. 
In short, the minimal CP4-based construction presented here can be used as a primer for introduction of additional features and
deriving their phenomenological and astroparticle consequences.

\bigskip

{\bf Acknowledgments.}
I am grateful to Ivo de Medeiros Varzielas, Alejandro Ibarra, Luis Lavoura, and Avelino Vicente for many valuable comments. 
My work was supported by the Portuguese
\textit{Fun\-da\-\c{c}\~{a}o para a Ci\^{e}ncia e a Tecnologia} (FCT) through the Investigator contract IF/00989/2014/CP1214/CT0004
under the IF2014 Program and in part by contracts UID/FIS/00777/2013 and CERN/FIS-NUC/0010/2015,
which are partially funded through POCTI, COMPETE, QREN, and the European Union. 
I also acknowledge the support from National Science Center, Poland, via the project Harmonia (UMO-2015/18/M/ST2/00518).


\begin{thebibliography}{99}

\bibitem{King:2017guk} 
  S.~F.~King,
  Prog.\ Part.\ Nucl.\ Phys.\  {\bf 94}, 217 (2017)
  [arXiv:1701.04413 [hep-ph]].


\bibitem{Cai:2017jrq} 
  Y.~Cai, J.~Herrero-García, M.~A.~Schmidt, A.~Vicente and R.~R.~Volkas,
  arXiv:1706.08524 [hep-ph].


\bibitem{King:2014nza} 
  S.~F.~King, A.~Merle, S.~Morisi, Y.~Shimizu and M.~Tanimoto,
  New J.\ Phys.\  {\bf 16}, 045018 (2014)
  [arXiv:1402.4271 [hep-ph]].


\bibitem{Altarelli:2010gt} 
  G.~Altarelli and F.~Feruglio,
  Rev.\ Mod.\ Phys.\  {\bf 82}, 2701 (2010)
  [arXiv:1002.0211 [hep-ph]].


\bibitem{Ma:2006km} 
  E.~Ma,
  Phys.\ Rev.\ D {\bf 73}, 077301 (2006)
  [hep-ph/0601225].


\bibitem{Ho:2013hia} 
  S.~Y.~Ho and J.~Tandean,
  Phys.\ Rev.\ D {\bf 87}, 095015 (2013)
  [arXiv:1303.5700 [hep-ph]].


\bibitem{Hessler:2016kwm} 
  A.~G.~Hessler, A.~Ibarra, E.~Molinaro and S.~Vogl,
  JHEP {\bf 1701}, 100 (2017)
  [arXiv:1611.09540 [hep-ph]].


\bibitem{Toma:2013zsa} 
  T.~Toma and A.~Vicente,
  JHEP {\bf 1401}, 160 (2014)
  [arXiv:1312.2840 [hep-ph]].


\bibitem{Vicente:2014wga} 
  A.~Vicente and C.~E.~Yaguna,
  JHEP {\bf 1502}, 144 (2015)
  [arXiv:1412.2545 [hep-ph]].


\bibitem{Barbieri:2006dq} 
  R.~Barbieri, L.~J.~Hall and V.~S.~Rychkov,
  Phys.\ Rev.\ D {\bf 74}, 015007 (2006)
  [hep-ph/0603188].


\bibitem{LopezHonorez:2006gr} 
  L.~Lopez Honorez, E.~Nezri, J.~F.~Oliver and M.~H.~G.~Tytgat,
  JCAP {\bf 0702}, 028 (2007)
  [hep-ph/0612275].


\bibitem{Ivanov:2017dad} 
  I.~P.~Ivanov,
  Prog.\ Part.\ Nucl.\ Phys.\  {\bf 95}, 160 (2017)
  [arXiv:1702.03776 [hep-ph]].


\bibitem{Keus:2014jha} 
  V.~Keus, S.~F.~King, S.~Moretti and D.~Sokolowska,
  JHEP {\bf 1411}, 016 (2014)
  [arXiv:1407.7859 [hep-ph]].


\bibitem{Bonilla:2014xba} 
  C.~Bonilla, D.~Sokolowska, N.~Darvishi, J.~L.~Diaz-Cruz and M.~Krawczyk,
  J.\ Phys.\ G {\bf 43}, no. 6, 065001 (2016)
  [arXiv:1412.8730 [hep-ph]].


\bibitem{Cordero-Cid:2016krd} 
  A.~Cordero-Cid, J.~Hernández-Sánchez, V.~Keus, S.~F.~King, S.~Moretti, D.~Rojas and D.~Sokołowska,
  JHEP {\bf 1612}, 014 (2016)
  [arXiv:1608.01673 [hep-ph]].


\bibitem{Restrepo:2013aga} 
  D.~Restrepo, O.~Zapata and C.~E.~Yaguna,
  JHEP {\bf 1311}, 011 (2013)
  [arXiv:1308.3655 [hep-ph]].


\bibitem{Simoes:2017kqb} 
  C.~Simoes and D.~Wegman,
  JHEP {\bf 1704}, 148 (2017)
  [arXiv:1702.04759 [hep-ph]].


\bibitem{Farzan:2012ev} 
  Y.~Farzan, S.~Pascoli and M.~A.~Schmidt,
  JHEP {\bf 1303}, 107 (2013)
  [arXiv:1208.2732 [hep-ph]].


\bibitem{Ivanov:2015mwl} 
  I.~P.~Ivanov and J.~P.~Silva,
  Phys.\ Rev.\ D {\bf 93}, no. 9, 095014 (2016)
  [arXiv:1512.09276 [hep-ph]].


\bibitem{Ivanov:2011ae} 
  I.~P.~Ivanov, V.~Keus and E.~Vdovin,
  J.\ Phys.\ A {\bf 45}, 215201 (2012)
  [arXiv:1112.1660 [math-ph]].


\bibitem{Aranda:2016qmp} 
  A.~Aranda, I.~P.~Ivanov and E.~Jiménez,
  Phys.\ Rev.\ D {\bf 95}, no. 5, 055010 (2017)
  [arXiv:1608.08922 [hep-ph]].


\bibitem{Ferreira:2017tvy} 
  P.~M.~Ferreira, I.~P.~Ivanov, E.~Jiménez, R.~Pasechnik and H.~Serôdio,
  arXiv:1711.02042 [hep-ph].


\bibitem{Ecker:1987qp} 
  G.~Ecker, W.~Grimus and H.~Neufeld,
  J.\ Phys.\ A {\bf 20}, L807 (1987).


\bibitem{Weinberg:1995mt} 
  S.~Weinberg,
  ``The Quantum theory of fields. Vol. 1: Foundations,''
  Cambridge University Press (1995)


\bibitem{Grimus:1999wm} 
  W.~Grimus and H.~Neufeld,
  Phys.\ Lett.\ B {\bf 486}, 385 (2000)
  doi:10.1016/S0370-2693(00)00769-3
  [hep-ph/9911465].


\bibitem{Hehn:2012kz} 
  D.~Hehn and A.~Ibarra,
  Phys.\ Lett.\ B {\bf 718}, 988 (2013)
  doi:10.1016/j.physletb.2012.11.034
  [arXiv:1208.3162 [hep-ph]].


\bibitem{Ferreira:2016sbb} 
  P.~M.~Ferreira, W.~Grimus, D.~Jurciukonis and L.~Lavoura,
  JHEP {\bf 1607}, 010 (2016)
  doi:10.1007/JHEP07(2016)010
  [arXiv:1604.07777 [hep-ph]].


\bibitem{Hirsch:2004he} 
  M.~Hirsch and J.~W.~F.~Valle,
  New J.\ Phys.\  {\bf 6}, 76 (2004)
  doi:10.1088/1367-2630/6/1/076
  [hep-ph/0405015].


\bibitem{Ibarra:2011gn} 
  A.~Ibarra and C.~Simonetto,
  JHEP {\bf 1111}, 022 (2011)
  doi:10.1007/JHEP11(2011)022
  [arXiv:1107.2386 [hep-ph]].


\bibitem{Wegman:2017zui} 
  D.~Wegman,
  arXiv:1711.08004 [hep-ph].

\end{thebibliography}
\end{document}